\documentclass[showpacs,preprintnumbers,amsmath,amssymb,floatfix,prd,11pt]{revtex4}
\usepackage[utf8]{inputenc}
\usepackage{epsfig}
\usepackage{amsmath}
\usepackage{subfig}
\usepackage{ulem}
\usepackage[dvipsnames]{xcolor}

\begin{document}
\title{The full set of Polarized Deep Inelastic Scattering Structure Functions at NNLO accuracy} 
\author{Ignacio Borsa}  
\email{iborsa@df.uba.ar}
\affiliation{Universidad de Buenos Aires and IFIBA, Facultad de Ciencias Exactas y  
Naturales, Departamento de Física. Buenos Aires, Argentina.}
\author{Daniel de Florian}  
\email{deflo@unsam.edu.ar}
\author{Iv\'an Pedron}  
\email{ipedron@unsam.edu.ar}
\affiliation{International Center for Advanced Studies (ICAS), ICIFI and ECyT-UNSAM, 25 de Mayo y Francia, (1650) Buenos Aires, Argentina}

\begin{abstract}

We  present the second order contributions to the coefficient functions for the parity violating polarized structure functions $g_L$ and $g_4$, thus completing the $\mathcal{O}(\alpha_S^2)$ knowledge on DIS structure functions. We obtain the missing $\mathcal{O}(\alpha_S^2)$ pieces from the known parity conserving unpolarized coefficient functions. We also present a phenomenological analysis for the phase space region the future Electron-Ion Collider is set to explore. 

\end{abstract}

\maketitle

\section{Introduction}

The deep inelastic scattering (DIS) process plays a fundamental role in the analysis of polarized QCD phenomena, and ultimately in our understanding of the spin structure of the proton in terms of polarized parton distributions functions (pPDFs). In addition to the pure QED contributions due to photon exchange, which have been thoroughly studied in pPDFs global analyses \cite{deFlorian:2008mr, Gluck:2000dy, DeFlorian:2019xxt,Ethier:2017zbq,Nocera:2014gqa}, the DIS process also receives contributions associated to the exchange of the weak $Z$ and $W^{\pm}$ bosons, which introduce parity violating terms to the cross section. Both neutral and charged current (NC and CC, respectively) DIS involve quark PDF combinations different from those of the pure photon counterpart, making the polarized DIS data an important source of complementary information on the proton spin decomposition, since it allows to disentangle the individual quark contributions from its corresponding antiquark ones~\cite{Aschenauer:2013iia, Boer:2011fh}. Nonetheless, and contrary to the unpolarized case, the rather low values of $Q^2$ explored by polarized DIS experiments so far, for which weak contributions are highly suppressed, have made the study of $Z$ and $W$-mediated processes rather unnecessary. 

However, with the construction of the new Electron-Ion-Collider (EIC) on the horizon, there is a prospect of reaching an unprecedented level of precision in measurements of polarized processes and to extend the kinematical coverage in terms of the DIS variables $x$ and $Q^2$~\cite{Accardi:2012qut}. Besides the extension of the kinematical range, the EIC will give access to the hadron helicity states independently, and thus allow to measure the asymmetries with polarized protons and unpolarized leptons for the first time \cite{Boer:2011fh}. These new precision measurement need to be matched by correspondingly accurate theoretical predictions. As it is already the standard for the unpolarized sector in Large-Hadron-Collider (LHC) computations, next-to-next-to-leading order (NNLO) calculations are becoming a benchmark for polarized processes, with results already available for inclusive process, such as Drell-Yan~\cite{Ravindran:2003gi} and pure QED DIS~\cite{Zijlstra:1993sh}, the helicity splitting functions~\cite{Vogt:2008yw, Moch:2014sna, Moch:2015usa}, as well as the recent addition of exclusive process like jet production in DIS~\cite{Borsa:2020ulb,Borsa:2020yxh}, $W$ boson production in proton-proton collisions~\cite{Boughezal:2021wjw} and semi-inclusive DIS (in an approximated form) \cite{Abele:2021nyo,Abele:2022wuy}. In this paper, we present the $\mathcal{O}(\alpha_S^2)$ parity violating (longitudinally) polarized structure functions $g_4$ and $g_L$, in order to match the NNLO precision for inclusive NC and CC DIS.

Our paper is organized as follows: in section~\ref{sec:sf} we introduce and provide the expressions of the polarized DIS structure functions at $\mathcal{O}(\alpha_S^2)$. In section~\ref{sec:single-jets} we analyze the impact of the second order corrections to the polarized structure functions, for both NC and CC processes and we discuss the combination of polarized structure functions related to the single-spin cross section. Finally, in section~\ref{sec:conclusion} we summarize our work and present our conclusions.

\section{Polarized structure functions} \label{sec:sf}

\begin{figure}
 \epsfig{figure= 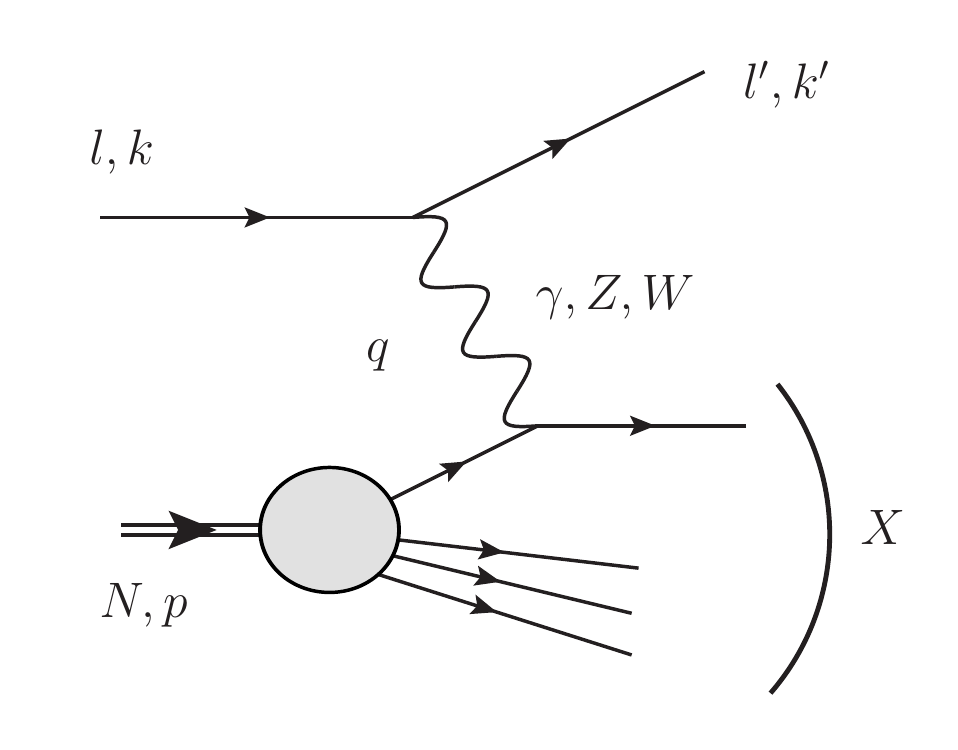, width=0.48\textwidth}
  \caption{Leading order diagram for the lepton-nucleon deep-inelastic scattering. The exchanged particle can be either a photon $\gamma$ or a weak boson $Z,\,W^{\pm}$, and carries momentum $q$.}
  \label{fig_DIS}
\end{figure}

In the following we consider the inclusive lepton-nucleon DIS process, defined as $$l(k)+N(p)\rightarrow l'(k')+X,$$ where, for simplicity, we assume the incoming lepton is either an electron or a positron and work within the lowest order approximation in the EW theory, i.e. assuming single-boson exchange. Here, $k$ and $p$ are the momenta of the incoming lepton and nucleon, respectively, $k'$ is the momentum of the outgoing scattered lepton (either an electron/positron in the NC or a neutrino/antineutrino in the CC case, respectively), and $X$ represents the whole recoiling hadronic final state. The particle exchanged in the process can be any of the electroweak bosons $\gamma$, $Z$ or $W^{\pm}$, and carries a momentum $q=k-k'$ determined by the lepton kinematics. The usual variables utilized in the description of DIS are the boson virtuality $Q^{2}$, the Bjorken variable $x$ and the inelasticity $y$, which are defined by
\begin{equation}
    Q^{2}=-q^{2}, \qquad x=\frac{Q^{2}}{2 p\cdot q}, \qquad y=\frac{q\cdot p}{k\cdot p}.
\end{equation}

\noindent While in the CC case the kinematics of the outgoing neutrino may not be experimentally accessible, the values of $x$ and $Q^2$ can be reconstructed from the hadronic final state using the Jacquet-Blondel method \cite{Aschenauer:2013iia}. The diagram for the lowest order contribution is shown in Fig.~\ref{fig_DIS}.

Following the notation from \cite{Workman:2022ynf}, the DIS cross section can be expressed in terms of the product of a hadronic ($W$) and a leptonic ($L$) tensor
\begin{equation}
   \frac{d^2 \sigma}{dx dy} = \frac{2 \pi y \alpha^2}{Q^4} \sum_i \eta_i L_i^{\mu \nu} W_{\mu \nu}^i,
\label{eq_tensors}
\end{equation}

\noindent with the summation over $i$ indicating the contributions associated to the different gauge bosons. For NC processes, the cross section receives contributions from the exchange of $\gamma$, $Z$ as well as their interference, so $i = \gamma, Z$, $\gamma / Z$. For CC processes, $i = W$. The factors $\eta_i$ are the ratios of the corresponding propagators and couplings to the ones of the photon exchange
\begin{equation}
\begin{split}
    \eta_{\gamma} &= 1, \quad 
    \eta_{\gamma / Z} = \left(\frac{G_F M_Z^2}{2 \sqrt{2} \pi \alpha}\right) \left( \frac{Q^2}{Q^2+M_Z^2}\right), \\
    \eta_{Z} &= \eta_{\gamma / Z}^2, \quad 
    \eta_{W} = \frac{1}{2} \left(\frac{G_F M_W^2}{4 \pi \alpha} \frac{Q^2}{Q^2+M_W^2}\right)^2.
\end{split}    
\end{equation}

The first piece of the cross section in Eq.~\eqref{eq_tensors} is the leptonic tensor $L^{\mu \nu}_i$. It corresponds to the square amplitude of the QED/EW interaction vertex between the boson and the leptons. In terms of the charge $e$ and helicity $\lambda$ of the incoming massless lepton (with $\lambda^2=1$), it has the following expressions for each boson contribution:
\begin{equation}
\begin{split}
    L^{\mu \nu}_{\gamma} &= 2 \left( - k \cdot k' g^{\mu \nu}+ k^{\mu} k'^{\nu} + k'^{\mu} k^{\nu} - i \lambda \epsilon^{\mu \nu \alpha \beta} k_{\alpha} k'_{\beta} \right), \\
    L^{\mu \nu}_{Z} &= \left( g_V^e + e \lambda g_A^e \right)^2 L^{\mu \nu}_{\gamma}, \\
    L^{\mu \nu}_{\gamma/Z} &= \left( g_V^e + e \lambda g_A^e \right) L^{\mu \nu}_{\gamma}, \\
    L^{\mu \nu}_{W} &= \left( 1 + e \lambda \right)^2 L^{\mu \nu}_{\gamma},
\end{split}
\label{eq_lepten}
\end{equation}

\noindent where $g_V^e = -\frac{1}{2}+2 \sin^2 \theta_W$ and $g_A= -\frac{1}{2}$, and $e=\pm1$ is the charge of the incoming lepton. In the case of pure QED, the leptonic tensor has a symmetric and an antisymmetric part, with the latter being proportional to the lepton helicity. The case of the weak bosons is more involved, since the axial coupling mixes the helicity dependence in both structures. However, the key point to notice is that the parity conserving terms proportional to the helicity have exactly the same tensor structure as the parity violating terms independent of $\lambda$, and vice-versa.

The other piece in Eq.~\eqref{eq_tensors} is the hadronic tensor $W^i_{\mu \nu}$, which includes the interaction of the EW current corresponding to the boson of type $i$ with the nucleon. In inclusive DIS, and up to twist-two terms, it is described in terms of the (polarized) unpolarized structure functions ($g$) $F$, defined as
\begin{equation}
\begin{split}
    W^i_{\mu \nu} = & \left( - g_{\mu \nu} + \frac{q_{\mu} q_{\nu}}{q^2} \right) \left[ F_1^i(x,Q^2)- \frac{h}{2} \ g_5^i(x,Q^2) \right] \\
    & + \frac{\left(p_{\mu}-\frac{p \cdot q}{q^2} q_{\mu}\right) \left(p_{\nu}-\frac{p \cdot q}{q^2} q_{\nu}\right)}{p \cdot q} \left[ F_2^i(x,Q^2)- \frac{h}{2} \  g_4^i(x,Q^2) \right] \\
    & - i \epsilon_{\mu \nu \alpha \beta} \frac{q^{\alpha} p^{\beta}}{2 p \cdot q} \left[ F_3^i(x,Q^2)+ h \ g_1^i(x,Q^2) \right],
\end{split}
\label{eq_hadronic_tensor}
\end{equation}

\noindent where $h$ is the helicity of the incoming hadron. The $g_2$ and $g_3$ contributions are excluded since they are suppressed by powers of $M^2/Q^2$, with $M$ being the hadron mass, in processes with longitudinal hadron polarization \cite{Forte:2001ph}. Other terms proportional to $M^2/Q^2$ accompanying the $g$'s structure functions were also neglected.

The total unpolarized and the double (longitudinally) polarized DIS cross sections for $j=\mathrm{NC},\mathrm{CC}$ are defined by the helicity combinations
\begin{equation}
    \begin{split}
        \frac{d^2 \sigma^j}{dx dy}&=\frac{1}{4}\sum_{\lambda,h=\pm1}\frac{d^2 \sigma^j(\lambda,h)}{dx dy},\\
        \frac{d^2 \Delta \sigma^j}{dx dy}&=\frac{1}{4}\sum_{\lambda,h=\pm1}\,\lambda\,h\,\frac{d^2 \sigma^j(\lambda,h)}{dx dy},
    \end{split}
\end{equation}

\noindent and they can be expressed in terms of the structure function using Eqs.~(\ref{eq_tensors},~\ref{eq_lepten},~\ref{eq_hadronic_tensor}) as:
\begin{equation}
\begin{split}
    \frac{d^2 \sigma^j}{dx dy} = & \frac{4 \pi \alpha^2}{x y Q^2} \xi^j \left\{ \left[ 1 + (1-y)^2 \right] F_2^j - y^2 F_L^j \mp x \left[ 1 - (1-y)^2 \right] F_3^j \right\}, \\
    \frac{d^2 \Delta \sigma^j}{dx dy} = & \frac{4 \pi \alpha^2}{x y Q^2} \xi^j \left\{ -\left[ 1 + (1-y)^2 \right] g_4^j + y^2 g_L^j \mp 2 x \left[ 1 - (1-y)^2 \right] g_1^j \right\},
\end{split}
\label{eq_totalDIS}
\end{equation}

\noindent with $\mp$ for positive or negatively charged leptons $l^{\pm}$, and the longitudinal structure functions defined as usual by
\begin{equation}
\begin{split}
    F_L^j = F_2^j - 2 x F_1^j, \\
    g_L^j = g_4^j - 2 x g_5^j.
\end{split}
\end{equation}

The variable $\xi$ takes the values $\xi^{\mathrm{NC}}=1$ and $\xi^{\mathrm{CC}}=2$, while the NC and CC structure functions correspond to the combinations

\begin{equation}\label{eq:couplings_SF}
    \begin{split}
        F_a^{\mathrm{CC}}&=F_a^{\mathrm{W}}\qquad a=1,2,3, \\
        F_a^{\mathrm{NC}}&= F_a^{\gamma}- g_V^e\,\eta_{\gamma / Z}\,F_a^{\gamma / Z}+({g_V^e}^2+{g_A^e}^2)\, \eta_Z\,F_a^Z\qquad a=2,L, \\
        F_3^{\mathrm{NC}}&= F_3^{\gamma}- g_A^e\,\eta_{\gamma / Z}\,F_3^{\gamma / Z}+2 g_V^e g_A^e\, \eta_Z\,F_3^Z,\\
        g_a^{\mathrm{CC}}&=g_a^{\mathrm{W}}\qquad a=1,2,3, \\
        g_1^{\mathrm{NC}}&= g_1^{\gamma}- g_V^e\,\eta_{\gamma / Z}\,g_1^{\gamma / Z}+({g_V^e}^2+{g_A^e}^2)\, \eta_Z\,g_1^Z,\\
        g_a^{\mathrm{NC}}&= g_A^e\,\eta_{\gamma / Z}\,g_a^{\gamma / Z}-2 g_V^e g_A^e\, \eta_Z\,g_a^Z\qquad a= 4,L.
    \end{split}
\end{equation}

Due to factorization, all the long-distances effects associated to the initial state nucleon can be separated from the hard scattering, and the structure functions $F^j_{a=2,3,L}$ and $g^j_{a=1,4,L}$ can be written in terms of convolutions between perturbatively calculable coefficient functions and parton distribution functions $f$ and $\Delta f$ (PDFs),
\begin{equation}\label{eq_fact_xsection}
    F^j_a=\sum_f C^{j,f}_a\otimes f
    ,\qquad g^j_a=\sum_f \Delta C^{j,f}_a\otimes \Delta f
    ,
\end{equation}

\noindent with $\otimes$ denoting the usual convolution, and each of the coefficient functions is calculated as an expansion in $\alpha_S$
\begin{equation}
(\Delta)C^{j, f}_a = \ (\Delta)C^{j, f (0)}_a  + \frac{\alpha_S}{4 \pi} \ (\Delta)C^{j, f (1)}_a + \left( \frac{\alpha_S}{4 \pi} \right)^2 (\Delta)C^{j, f (2)}_a + ...
\end{equation}

At the LO in QCD, the coefficients are fairly simple ($\ (\Delta)C^{j, f (0)}_a (z) = \delta(1-z)$ for $a\neq L$ and $\ (\Delta)C^{j, f (0)}_L (z) = 0$ ) and the structure functions are  given by different combinations of quark PDFs, as in the naive parton model. Considering the process where the initial lepton is an electron, the combinations for the NC case are

\begin{equation}
\begin{split}
    \left[ F_2^{\gamma},\ F_2^{\gamma/Z},\ F_2^{Z} \right] =& x \sum_q  \left[ e_q^2,\ 2 e_q g_V^q,\ {g_V^q}^2 + {g_A^q}^2 \right] (q+\bar{q}), \\
    \left[ F_3^{\gamma},\ F_3^{\gamma/Z},\ F_3^{Z} \right] =& \sum_q  \left[ 0,\ 2 e_q g_A^q,\ 2 g_V^q g_A^q \right] (q-\bar{q}), \\
    \left[ g_1^{\gamma},\ g_1^{\gamma/Z},\ g_1^{Z} \right] =& \frac{1}{2} \sum_q  \left[ e_q^2,\ 2 e_q g_V^q,\ {g_V^q}^2 + {g_A^q}^2 \right] (\Delta q+\Delta \bar{q}), \\
    \left[ g_4^{\gamma/Z},\ g_4^{Z} \right] =& x \sum_q  \left[ 2 e_q g_A^q,\ 2 g_V^q g_A^q \right] (\Delta q-\Delta \bar{q}),
\end{split}
\end{equation}

\noindent where in this case the quark coupling factors are $g_V^q = \pm \frac{1}{2} - 2 e_q \sin^2 \theta_W$ and $g_A^q = \pm \frac{1}{2}$, with $\pm$ according to whether $q$ is an up-type or down-type quark, respectively. The longitudinal structure functions are null at the lowest order, in accordance with the Callan-Gross and Dicus relations \cite{Callan:1969uq, Dicus:1972pq}. While this makes the $\mathcal{O}(\alpha_S^2)$ formally NLO for the longitudinal coefficients, in what follows we will refer to the $\mathcal{O}(\alpha_S^2)$ corrections as NNLO, having in mind that the NNLO DIS cross section receives contributions of that order. That is, the perturbative orders of $g_L$ are counted as those of $g_{1,4}$.

For the CC process, the structure functions at LO are given by the combinations
\begin{equation}
\begin{split}
    F_2^{W^-} =& 2 x \left[ \sum_{f<} (\bar{d} |V_{fd}|^2 + \bar{s} |V_{fs}|^2 + ...) + \sum_{f'<} ( u |V_{uf'}|^2 + c |V_{cf'}|^2 + ...) \right], \\
    F_3^{W^-} =& 2 \left[ -\sum_{f<} (\bar{d} |V_{fd}|^2 + \bar{s} |V_{fs}|^2 + ...) + \sum_{f'<} ( u |V_{uf'}|^2 + c |V_{cf'}|^2 + ...) \right], \\
    g_1^{W^-} =& \left[ \sum_{f<} (\Delta \bar{d} |V_{fd}|^2 + \Delta \bar{s} |V_{fs}|^2 + ...) + \sum_{f'<} ( \Delta u |V_{uf'}|^2 + \Delta c |V_{cf'}|^2 + ...) \right], \\
    g_4^{W^-} =& 2x \left[ \sum_{f<} (\Delta \bar{d} |V_{fd}|^2 + \Delta \bar{s} |V_{fs}|^2 + ...) - \sum_{f'<} ( \Delta u |V_{uf'}|^2 + \Delta c |V_{cf'}|^2 + ...) \right],
\end{split}
\end{equation}

\noindent where $V_{ff'}$ are the elements of the Cabibbo–Kobayashi–Maskawa (CKM) matrix, and $f<$ means that only active flavors below a certain threshold are kept. For the process with an initial positron, the structure functions are obtained via the flavor exchanges $d \longleftrightarrow u$, $s \longleftrightarrow c$, etc.

At higher perturbative orders, it is customary to recast Eq.~\eqref{eq_fact_xsection} in terms of the usual singlet (S) and nonsinglet (NS) quark PDF combinations, $\Sigma$ and $\Delta$,  as well as their corresponding coefficient functions. In the case of unpolarized DIS, the structure functions $F_a$ can then be expressed as

\begin{equation}
\begin{split}
    F^j_{a} (x, Q^2) =  \omega_{a} \int \frac{dz}{z} &\Bigg\{ \bigg(\frac{1}{N_f} \sum^{N_f}_{f=-N_f} \kappa^j_{f,a} \bigg) \left[ \Sigma \left(\frac{x}{z}, \mu_F^2\right) \ C^{j, \mathrm{S}}_a(z, Q^2/\mu_F^2) + G \left(\frac{x}{z}, \mu_F^2\right) \ C^{j, g}_a(z, Q^2/\mu_F^2) \right] \\
    & + \Delta \left(\frac{x}{z}, \mu_F^2\right) \ C^{j, \mathrm{NS}}_a(z, Q^2/\mu_F^2) \Bigg\}, \quad a=2,3,L \ ,
\label{eq_SF_coef}
\end{split}
\end{equation}

\noindent where $G$ denotes the gluon PDF, and the dependence on the renormalization scale is implicit in all quantities. The factor $\omega$ accounts for the different definitions of each structure function, with $\omega_{a}=x$ for $a=2,L$ and $\omega_{3}=1$. The singlet and nonsinglet combinations are, respectively
\begin{equation}\label{eq_S_NS}
    \begin{split}
        \Sigma(z,Q^2)&=\sum^{N_f}_{f=-N_f} \,q_f(z,Q^2),\\
        \Delta(z,Q^2)&=\sum^{N_f}_{f=-N_f} \kappa^j_{f,a} \ q_f(z,Q^2) - \left(\frac{1}{N_f} \sum^{N_f}_{k=-N_f} \kappa^j_{k,a} \right) \, \Sigma(z,Q^2).
    \end{split}
\end{equation}

\noindent The factors $\kappa^j_{f,a}$ in Eqs.~\eqref{eq_SF_coef} and \eqref{eq_S_NS} correspond to the products of couplings of the electron and the parton $f$ to the cross sections of the process $j$, and are given by

\begin{equation}\label{eq_kappa}
    \begin{aligned}[c]
    &\kappa_{f,2}^{\mathrm{NC}}=\kappa_{f,L}^{\mathrm{NC}}=e_f^2 - 2\,e_f\,g_V^f\,g_V^e\,\eta_{\gamma / Z}+  ({g_V^e}^2+{g_A^e}^2)({g_V^f}^2+{g_A^f}^2)\,\eta_{Z},\\
    &\kappa_{f,3}^{\mathrm{NC}}=4\,g_V^f\,g_A^f\,g_V^e\,g_A^e \,\eta_{\gamma / Z} - 2\,e_f\,g_A^f\,g_A^e\,\eta_{Z},\\
    &\kappa_{f,a}^{\mathrm{CC}}=\sum_{f'<} |V_{ff'}|^2, \qquad a=2,3,L.\\
    \end{aligned}
\end{equation}

\begin{figure}
 \epsfig{figure= 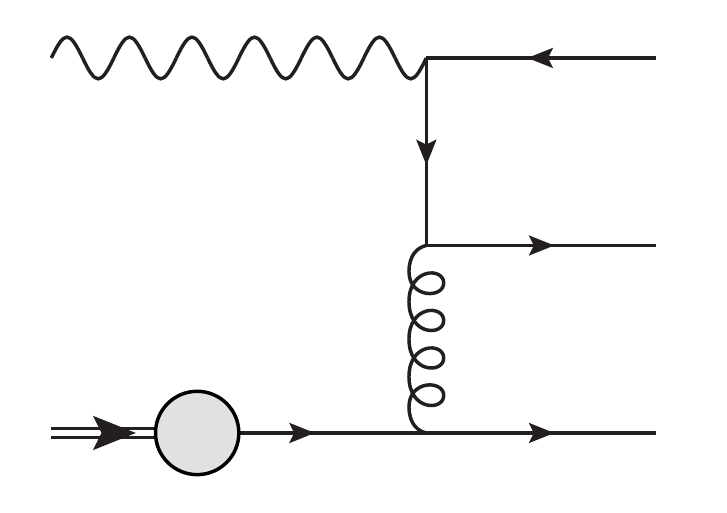, width=0.35\textwidth}
  \caption{Example of a Feynman diagram for the process $l+q\rightarrow l'+q+q'+\bar{q}'$ associated to purely-singlet contributions. The leptonic part of the diagram was omitted for simplicity.}
  \label{fig_diagram_PS}
\end{figure}

\noindent It should be noted that, in the case $j=\mathrm{CC}$, the summations in Eqs.~\eqref{eq_SF_coef} and in the definition of $\Delta$ are done only over the flavors that couple to the gauge boson. The usual decomposition of Eq.~\eqref{eq_SF_coef} in terms of singlet and non-singlet combinations of quarks is useful starting at NNLO, since it is from $\mathcal{O}(\alpha_S^2)$ that one gets $C_a^{j, \mathrm{S}}\neq C_a^{j,\mathrm{NS}}$ due to contributions from the partonic processes $l+q\rightarrow l'+q+q'+\bar{q}'$ opening at that order. The difference between the S and NS coefficients can be parameterized in terms of the purely-singlet (PS) contribution, $C_a^{j, \mathrm{PS}}=C_a^{j, \mathrm{S}}-C_a^{j, \mathrm{NS}}$, associated to processes as those depicted in Fig.~\ref{fig_diagram_PS}, in which the gauge boson couples to quarks of radiative origin. For the sake of the following discussion, it is actually useful to recast Eq.~\eqref{eq_SF_coef} in terms of the pure-singlet coefficient, as

\begin{equation}
\begin{split}
    F^j_{a} (x, Q^2) = \omega_{a} \int \frac{dz}{z} &\Bigg\{ \bigg(\frac{1}{N_f} \sum^{N_f}_{f=-N_f} \kappa^j_{f,a} \bigg) \left[ \Sigma \left(\frac{x}{z}, \mu_F^2\right) \, C^{j, \mathrm{PS}}_a(z, Q^2/\mu_F^2) + G \left(\frac{x}{z}, \mu_F^2\right) \, C^{j, g}_a(z, Q^2/\mu_F^2) \right] \\
    & + \sum^{N_f}_{j=-N_f} \kappa^j_{f,a}\,\, q_f\left(\frac{x}{z}, \mu_F^2\right) \ C^{j, \mathrm{NS}}_a(z, Q^2/\mu_F^2) \Bigg\}.
\label{eq_SF_coef_PS}
\end{split}
\end{equation} 

While in Eq.~\eqref{eq_SF_coef} and \eqref{eq_SF_coef_PS} we showed the unpolarized structure functions, a completely analogous expression stands for the polarized $g_{a=1,4,L}$ structure functions, in terms of polarized PDFs combinations and polarized coefficient functions. In that case, the constants $\Delta\kappa^j_{f,a}$, which 
can be obtained from those in Eq.~\eqref{eq_kappa}, and the $\Delta \omega_a$ are given by
\begin{equation}
\begin{split}
    \Delta \kappa^j_{f,1} &= \kappa^j_{f,2}, \quad 
    \Delta \omega_1 = \frac{1}{2}, \\
    \Delta\kappa^j_{f,4} &= \Delta\kappa^j_{f,L}=\kappa^j_{f,3}, \quad
    \Delta \omega_4 = \Delta \omega_L = x.
\end{split}    
\end{equation}
 
For unpolarized DIS, all the necessary ingredients to achieve $\mathcal{O}(\alpha_S^2)$ accuracy are known. Actually, the NNLO coefficients have already been available for a long time for $F_2$~\cite{Zijlstra:1992qd, vanNeerven:1991nn, Zijlstra:1991qc}, $F_L$~\cite{Zijlstra:1992qd, SanchezGuillen:1990iq} and $F_3$~\cite{Zijlstra:1992kj}, while NNLO extractions of PDFs are now standard. The picture for polarized DIS, on the other hand, is not as developed.  Besides the lack of NNLO polarized PDFs, only $g_1$ was calculated up to NNLO accuracy \cite{Zijlstra:1993sh}. Thus, the missing NNLO coefficient functions are those of the polarized parity violating structure functions $g_4$ and $g_L$, which have never been studied beyond the NLO level~\cite{deFlorian:1994wp,Stratmann:1995fn,Anselmino:1996cd,Forte:2001ph}. 
 
However, due to connections between parity violating polarized contributions and the parity conserving unpolarized ones, originated from the axial Ward identity, it is possible to obtain the missing coefficient functions from the already known ones in the massless limit. This kind of arguments were presented in our previous paper on dijet production \cite{Borsa:2021afb} as a useful tool to ease the implementation of polarized NC and CC processes. It goes as follows: since the gluon and pure singlet contributions (up to NNLO) vanish after integration over the final-state particles (due to charge conjugation arguments), and the triangle terms present in $Z$ boson exchange cancel when the complete weak isospin doublets are considered (that is, an even number of flavors), the remaining non-singlet coefficients of $g_4$ and $g_L$ can then be obtained from the non-singlet pieces of $F_2$ and $F_L$, respectively. This is a direct consequence of the axial Ward identity~\cite{Larin:1991tj}, and it can already be observed at the matrix element level of the cross section, since both the parity violating polarized and non parity violating unpolarized fermionic traces have the same structure, and can only differ up to the presence of an additional even number of $\gamma_5$ matrices, which cancel out when dealing properly with the HVBM scheme within dimensional regularization. This involves, for instance, the use of a symmetric definition of the axial vertex and an additional finite renormalization of the axial current, that are needed to effectively restore the anticommutativity of $\gamma^5$, and, therefore, the conservation of helicity in $d-$dimensions \cite{Larin:1993tq}. 
Specifically, one then gets
\begin{equation}
\begin{split}
    \Delta C_{4}^{j,g,(2)}&=\Delta C_{L}^{j,g,(2)}=0,\\
    \Delta C_4^{j,\text{NS},(2)}&=\Delta C_4^{j,\text{S},(2)}=C_2^{j,\text{NS},(2)},\\
    \Delta C_L^{j,\text{NS},(2)}&=\Delta C_L^{j,\text{S},(2)}=C_L^{j,\text{NS},(2)}.    
\end{split}    
\end{equation}

Similar arguments can be used to demonstrate the equivalence between the NS coefficients of the unpolarized parity violating structure function $F_3$ and the polarized parity conserving $g_1$, which was already noted in~\cite{Zijlstra:1993sh, Zijlstra:1993zs}.

\begin{figure}
 \epsfig{figure= 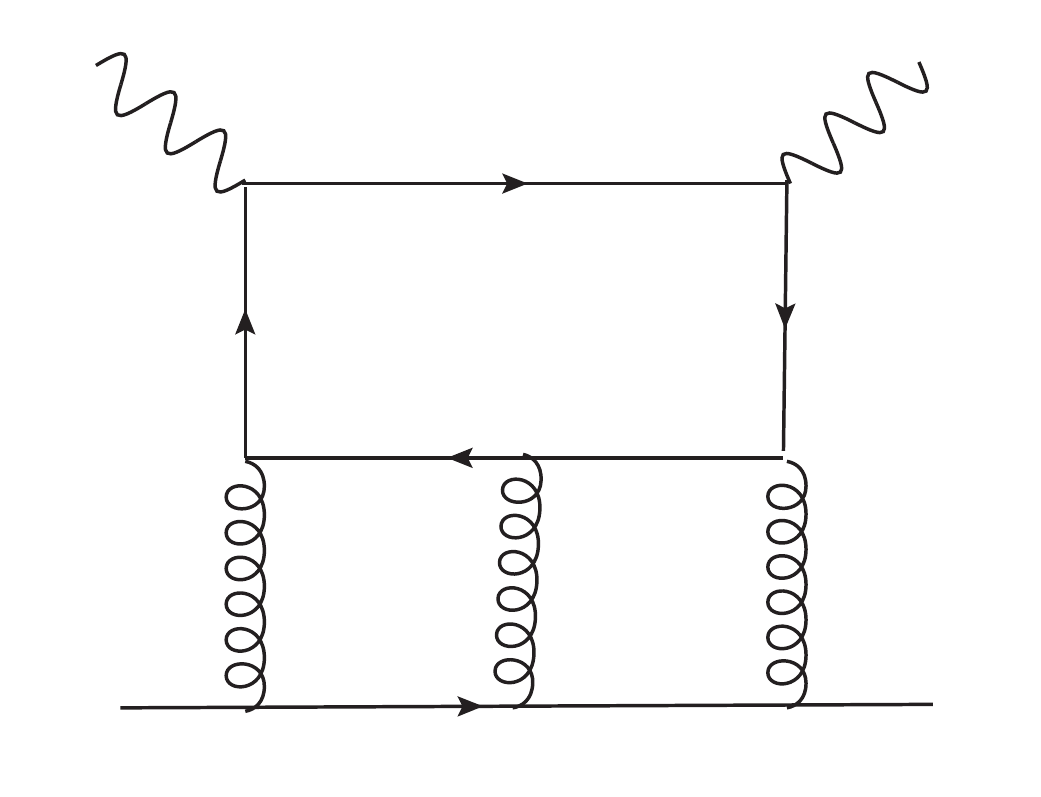, width=0.48\textwidth}
  \caption{Example of an N3LO trace that contributes to the polarized parity violating structure functions, that cannot be obtained from the unpolarized $F_2$ and $F_L$.}
  \label{fig_diagram}
\end{figure}

These relations between polarized and unpolarized structure functions are only valid up to $\mathcal{O}(\alpha_s^2)$. This is due to the new contributions showing up in the perturbative series at higher orders, which involve additional fermion loops that end up spoiling them~\cite{Zijlstra:1993zs}. In particular, traces like the one presented in Fig.~\ref{fig_diagram} are associated with pure singlet contributions that no longer vanish, and they need to be calculated from scratch, since they have a different trace structure from those in unpolarized processes. Note that at N3LO, those contributions like the one of Fig.~\ref{fig_diagram} are associated with a new color factor, in this case $(d^{a b c})^2$, and they can in principle be easily isolated.

Following these relations, we now complete the $\mathcal{O}(\alpha_s^2)$ knowledge of the missing polarized parity violating structure functions. The $\mathcal{O}(\alpha_s^2)$ non-singlet coefficients of $g_L$ and $g_4$ are simply obtained from those of the non-singlet ones of $F_L$ and $F_2$, respectively, which are given in Eqs.~(B.1) and (B.2) of ref.~\cite{Zijlstra:1992qd}. Regarding the singlet sector, the gluon contribution cancels out, and the quark singlet contribution is identical to the non-singlet one.

\section{Numerical Results of Polarized NNLO Proton Structure Functions}\label{sec:single-jets}

In this section we compute the polarized structure functions in order to evaluate the impact of the NNLO corrections. Since the EIC is set to reach high precision measurement of DIS observables, we focus on the case of polarized electron-proton collisions. We work at a fixed $Q^2 = 100$ GeV, and the renormalization and factorization scales are set at central values of $\mu_{F}^{2}=\mu_{R}^{2}=Q^2\equiv\mu_{0}^2$, with $\alpha_s$ evaluated at NLO accuracy with $\alpha_s(M_Z)=0.118$ and using $n_F=4$ active flavors. For the weak gauge bosons we use the masses $M_Z = 91.1876$ GeV and $M_W = 80.379$ GeV, with decay-widths of $\Gamma_Z = 2.4952$ GeV and $\Gamma_W = 2.085$ GeV. The electromagnetic coupling constant is set to $\alpha = 1/137$ and the Weinberg angle given by $\sin^2\theta_W = 0.23122$. The values used for the CKM matrix are $|V_{ud}|=0.9737$, $|V_{us}|=0.2245$, $|V_{ub}|=0.00382$, $|V_{cd}|=0.2210$, $|V_{cs}|=0.987$ and $|V_{cb}|=0.041$. 

For the polarized densities, we use the set from the latest DSSV global analysis \cite{deFlorian:2014yva,DeFlorian:2019xxt}. We note that, since there are no \textit{global} analyses of polarized PDFs available at NNLO (current extractions at that order are solely based on DIS data \cite{Taghavi-Shahri:2016idw}), we restrict ourselves to NLO PDFs \footnote{While this is not truly a fully consistent calculation at each perturbative order, it is particularly convenient in order to analyze the size of the NNLO corrections.}.

\begin{figure}
    \centering
    \subfloat{{\includegraphics[width=0.47\textwidth]{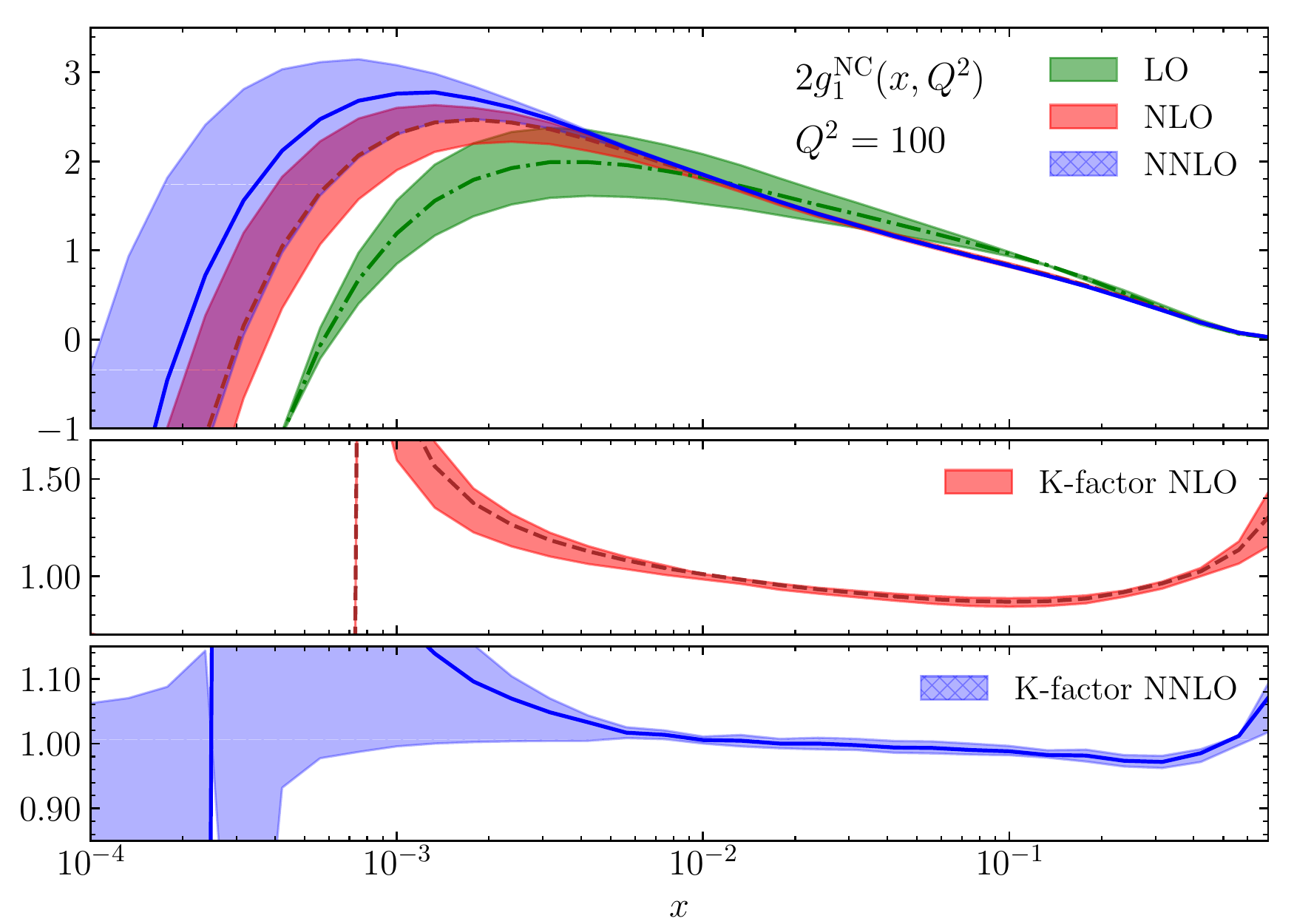} }}%
    \qquad
    \subfloat{{\includegraphics[width=0.47\textwidth]{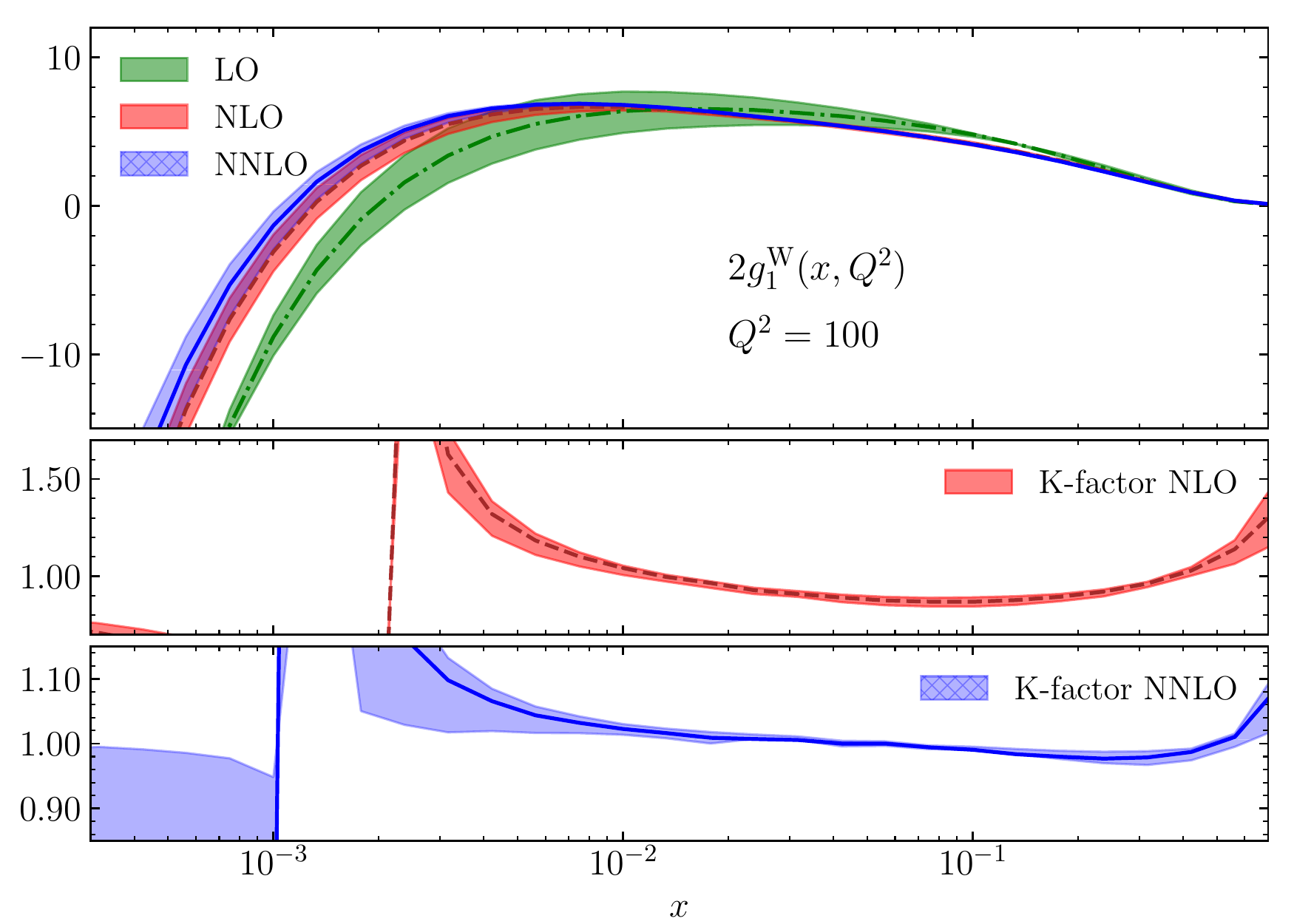} }}%
    \caption{Non parity violating structure function $2\, g_1$ for neutral (left) and charged current (right) exchange at $Q^2=100\,\mathrm{GeV}^2$, for the Leading (green), Next-to-Leading (red) and Next-to-Next-to-Leading Order (blue). The bands correspond to the theoretical uncertainty obtained by the independent variation of the renormalization and factorization scales. The lower insets in both cases represent the K-factors defined as the ratio of each order to the previous one.}%
    \label{fig:g1}%
\end{figure}

We begin by showing in Fig.~\ref{fig:g1} the already known parity conserving $g_1$ structure function as a function of $x$ for both NC and $W$ exchange. The bands at each successive order represent the theoretical uncertainty of the cross section, obtained by the independent 7-point variation of the renormalization and factorization scales ( $ \mu_0/2 \leq \mu_F,\mu_R \leq 2 \mu_0$ with $1/2 \leq \mu_F/\mu_R \leq 2$). The lower insets show the $K-$factors, i.e, the ratio of each perturbative order to the previous one. Higher order corrections are particularly large around $x \sim 10^{-3}$ due to the existence of a sign change. The overlap between the NNLO and NLO bands hints towards a convergence of the perturbative series. However, the scale bands can still be relatively big at lower values of $x$, particularly in the NC case.
It is also worth noticing that the LO band at low $x$ is artificially small. This is related to the fact that the gluon channel, with a contribution that is typically relevant in that region, only opens up at NLO. In DIS, as it happens in the case of hadronic-collisions at the LHC,  LO calculations fail to produce sensible results not only for the central values but also for the estimates of higher order uncertanties. 

\begin{figure}
    \centering
    \subfloat{{\includegraphics[width=0.47\textwidth]{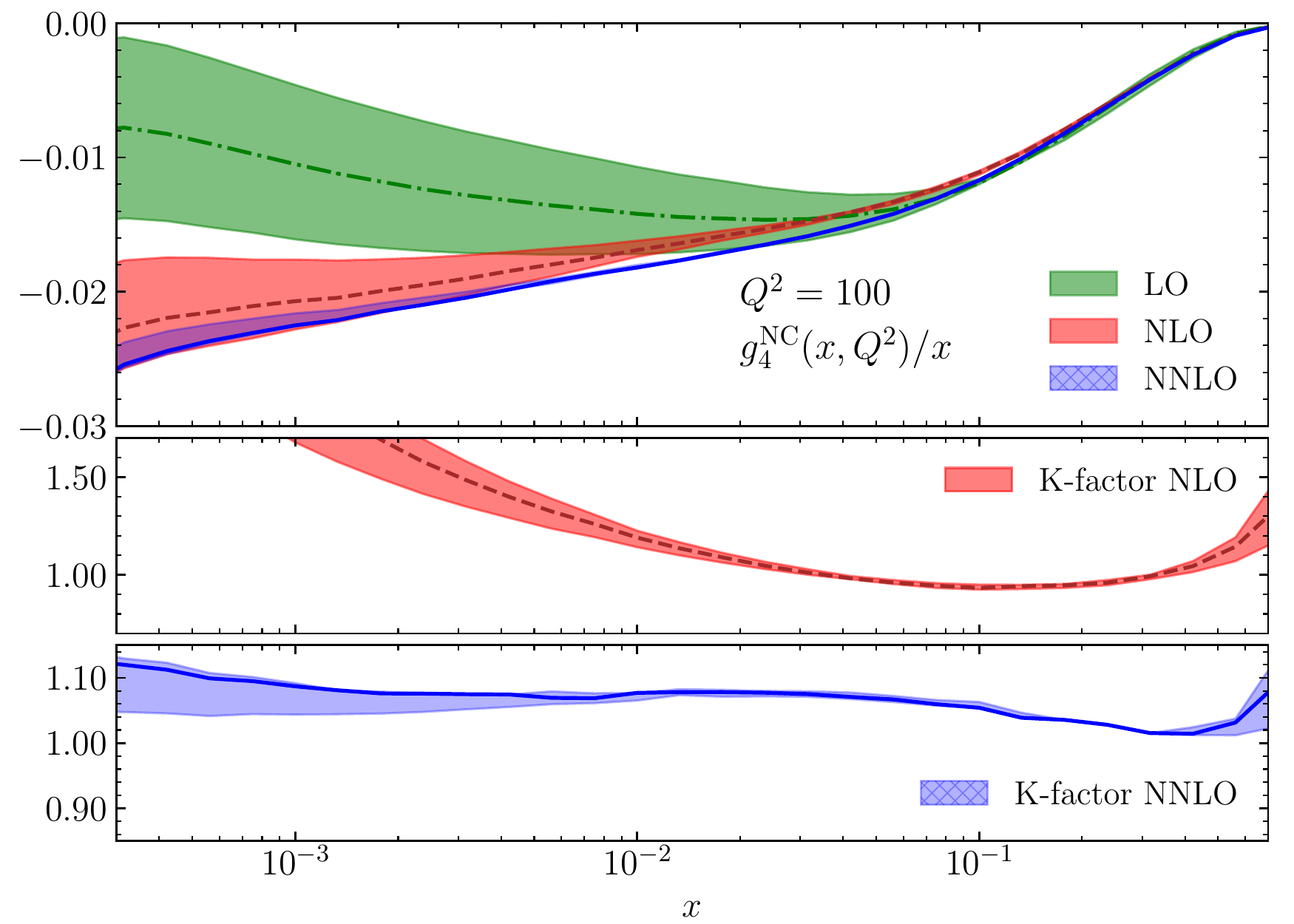} }}%
    \qquad
    \subfloat{{\includegraphics[width=0.47\textwidth]{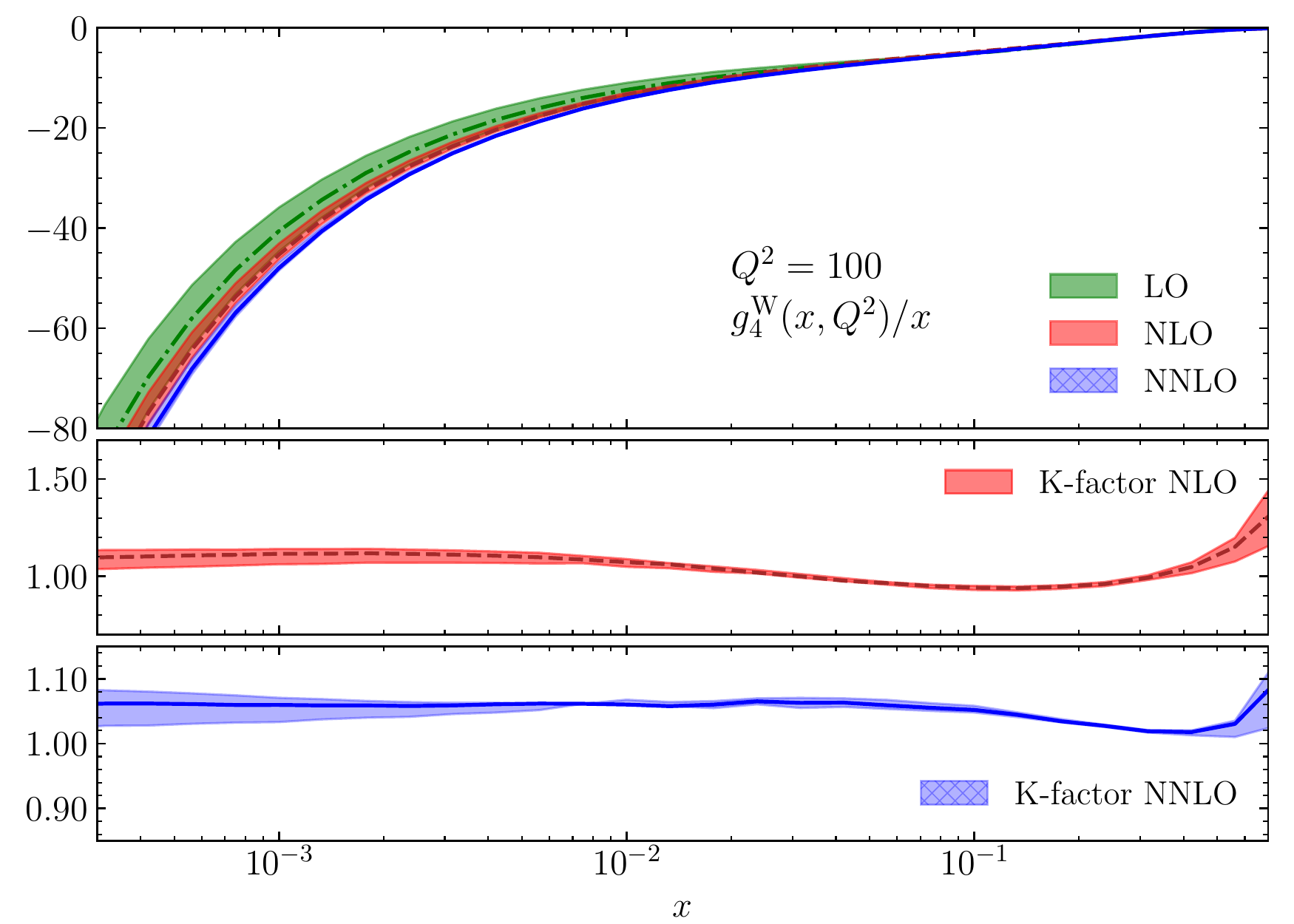} }}%
    \caption{Same as Fig.~\ref{fig:g1} for the parity violating structure function $g_4/x$.}%
    \label{fig:g4}%
\end{figure}

In Fig.~\ref{fig:g4} we present the first of the parity violating structure functions, $g_4$. In this case, even though there is a lack of overlap at higher values of $x$, there is a noticeable reduction in the NNLO and NLO bands, in both NC and $W$ exchange. The $K-$factors are greatly reduced at NNLO, stabilizing at around 10\% and 25\%, respectively, but they are still sizable. This highlights the importance of higher order corrections in polarized processes, even at $\mathcal{O}(\alpha_S^2)$. It is worth noticing that, in the NC case, there is no longer a pure photon contribution to the parity violating functions nor initial gluon or pure-singlet contribution, which accounts for the reduction in the uncertainty bands.

\begin{figure}
    \centering
    \subfloat{{\includegraphics[width=0.47\textwidth]{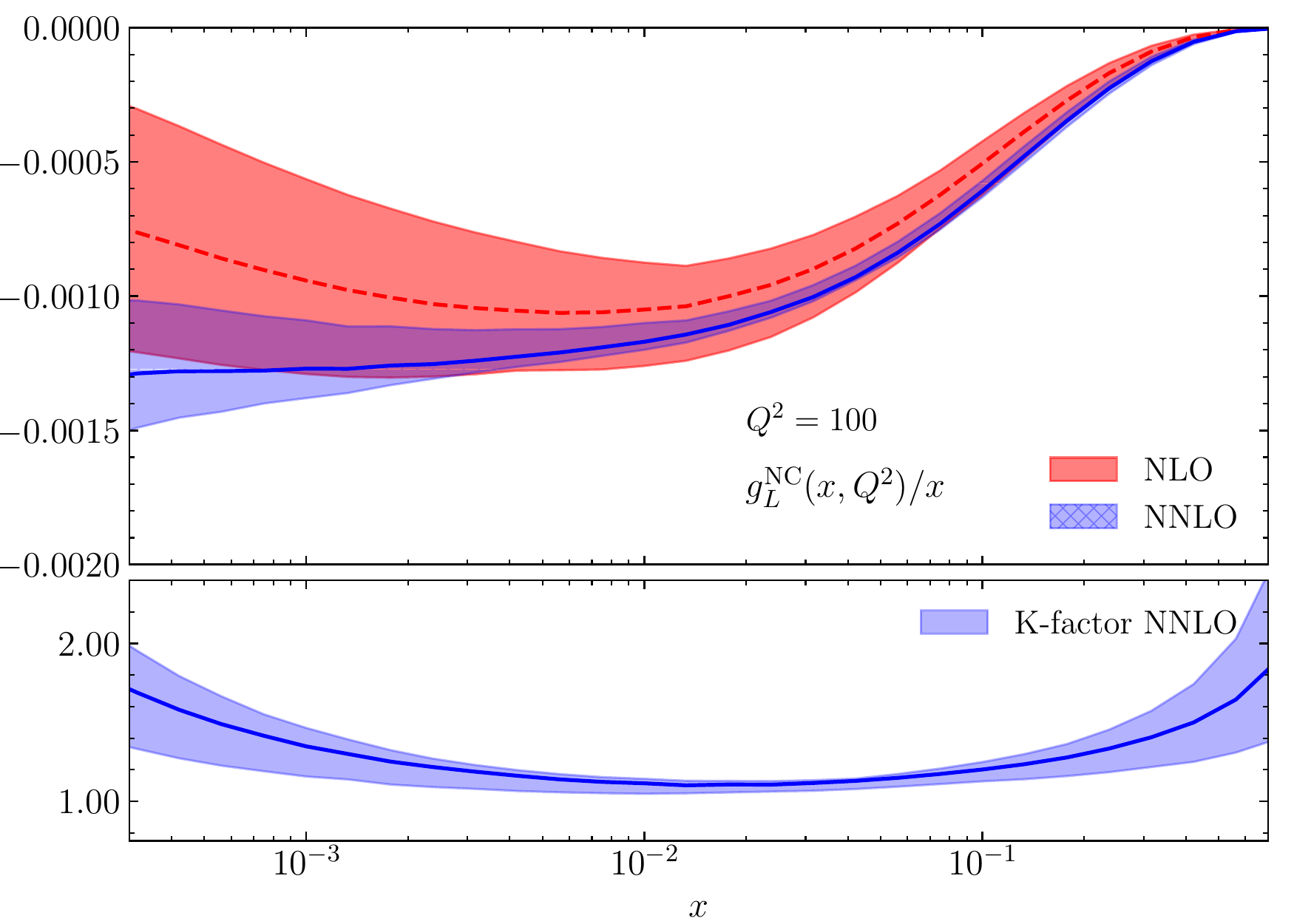} }}%
    \qquad
    \subfloat{{\includegraphics[width=0.47\textwidth]{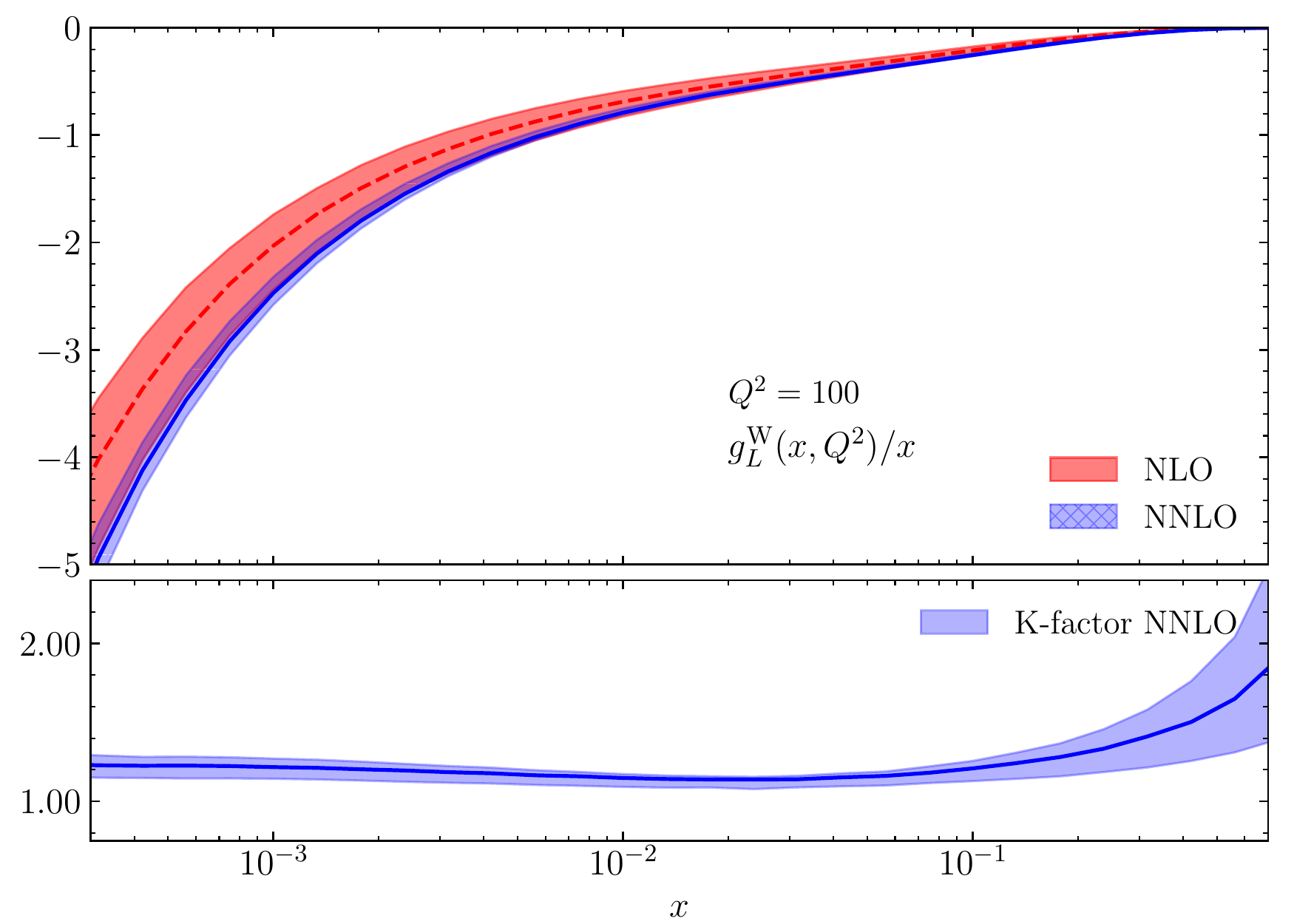} }}%
    \caption{Same as Fig.~\ref{fig:g1} for the parity violating structure function $g_L/x$. }%
    \label{fig:gL}%
\end{figure}

Similar comments can be made regarding the parity violating structure function $g_L$, which is presented in Fig.~\ref{fig:gL} in the same fashion as $g_1$ and $g_4$. Although there is overlap for most of the intermediate $x$ range, corrections are still sizable at high and low values of $x$, where they can reach as much as $75\%$. As mentioned, since $\Delta C_{L}^{i,f,(0)}=0$, the first non-vanishing contribution for the longitudinal structure function starts at $\mathcal{O}(\alpha_S)$.  While this makes the $\mathcal{O}(\alpha_S)$ formally LO, we still refer to it as NLO in the sense of it being a NLO contribution to the total DIS cross sections. 

Finally, we would like to call the attention to another possibly interesting combination  of polarized structure function to be observed at the EIC. Due to the parity violating structure of the EW interaction, it is possible to define {\it single-spin} structure functions that can still shed light on the polarized structure of the nucleon. The corresponding cross section can be obtained by considering the average over the spin of the lepton and taking the difference only over the helicity of the nucleon, as  
\begin{equation}
        \frac{d^2 \delta \sigma^j}{dx dy}=\frac{1}{4}\sum_{\lambda,h=\pm1}\,h\,\frac{d^2 \sigma^j(\lambda,h)}{dx dy}.
\end{equation}

In that case, the single-spin polarized structure functions are given by the same perturbative coefficients as for the double polarized case, but replacing the combinations of EW charges of Eq.~\eqref{eq:couplings_SF}. In the NC case, they are given by 
\begin{equation}\label{eq:couplings_SPSF}
    \begin{split}
        {g_1^{\mathrm{NC}}}_{\mathrm{SS}}&=  g_A^e\,\eta_{\gamma / Z}\,g_1^{\gamma / Z}-2 g_V^e g_A^e\, \eta_Z\,g_1^Z,\\
        {g_a^{\mathrm{NC}}}_{\mathrm{SS}}&= - g_V^e\,\eta_{\gamma / Z}\,g_a^{\gamma / Z}+({g_V^e}^2+{g_A^e}^2)\, \eta_Z\,g_a^Z\qquad a= 4,L.
    \end{split}
\end{equation}

\noindent  The use of the asymmetries for unpolarized leptons scattered off polarized nucleons is also discussed in \cite{Boer:2011fh}. In the CC case, since the coupling is purely chiral, only one of the leptonic polarization contributes to the process, and therefore the single-spin structure functions are analogous to the double polarized ones. 

\begin{figure}
    \centering
    \includegraphics[width=0.5\textwidth]{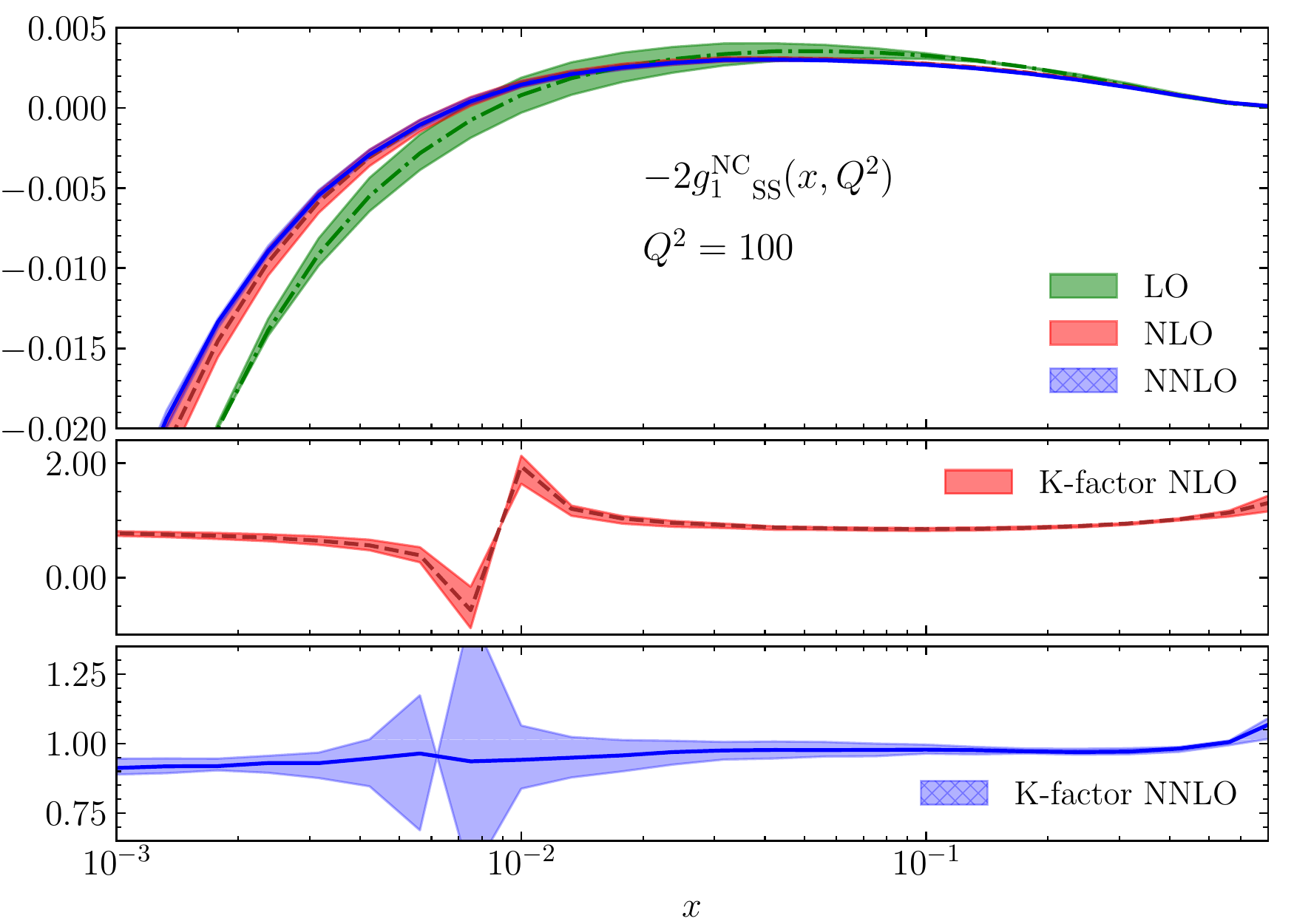}
    \caption{Single-spin structure function $-2\,{g_1^{\mathrm{NC}}}_{\mathrm{SS}}$ for NC exchange at $Q^2=100\,\mathrm{GeV}^2$, for the Leading (green), Next-to-Leading (red) and Next-to-Next-to-Leading Order (blue).}%
    \label{fig:g1ss}%
\end{figure}

Notice that the pure photon contribution in the NC structure function ${g_1^{\mathrm{NC}}}_{\mathrm{SS}}$ cancels trivially since the {\it QED component} of leptonic tensor is completely symmetric when the average over polarization is taken, while the same contribution of $g_1$ to the hadronic tensor is completely anti-symmetric. Therefore, single-spin observables provide an easy way of suppressing the photon contribution to NC process and a cleaner access to purely weak effects and provide access as a probe for new physics. As an example, in Fig.~\ref{fig:g1ss} we show the single-spin structure function $g_1$ for NC exchange at $Q^2=100\,\mathrm{GeV}^2$. As expected, since the perturbative coefficients are the same as for the double spin counterparts, the results show similar features regarding the stability of the higher order corrections.

On the other hand, the charges in ${g_a^{\mathrm{NC}}}_{\mathrm{SS}}$ imply a suppression of the single-spin $g_4$ and $g_L$ structure functions compared to the double spin ones. However, since $g_A^e$ is about one order of magnitude larger that $g_V^e$, the pure $Z$ exchange gets relatively enhanced compared to the $\gamma/Z$ interference.

\section{Conclusions}\label{sec:conclusion}

In this work we present the $\mathcal{O}(\alpha_S^2)$ corrections for the coefficient functions associated to the parity violating polarized structure functions $g_4$ and $g_L$, thus completing the $\mathcal{O}(\alpha_S^2)$ knowledge on DIS structure functions. We show that, since both the purely singlet and gluon contributions up to this order cancel, the missing coefficients can be obtained from the already known parity conserving unpolarized structure functions $F_2$ and $F_L$.

We also analyze the effects of the higher order corrections in the relevant kinematics for the future Electron-Ion Collider. In general, we observe a reduction of the $K$-factors and the scale dependence bands, pointing towards the stabilization of the perturbative series. This reduction is particularly noticeable in the case of $g_4$ and $g_L$, for which the $K$-factors are typically under $10\%$ with estimated theoretical uncertainties below $5\%$. 

Finally, we discuss the possibility of using combinations of \textit{single-spin} structure functions, considering only the hadron polarization. Since the single-spin structure functions include different combinations of EW charges, they provide an additional probe on the helicity structure of nucleons and a source of EW precision tests to be performed in the future EIC.

\acknowledgments

This work was partially supported by CONICET and ANPCyT.

\bibliography{refs}
\clearpage

\end{document}